\documentclass[11pt,a4paper]{article} % And comment these out for Sami's version
\pdfoutput=1 %
\usepackage{jcappub} %

\usepackage{amssymb}
\usepackage{graphics}
\usepackage{epsfig}
\usepackage{graphicx}
\usepackage{subfig}
\usepackage{bm}

\newcommand{\ud}{\mathrm{d}}

\newcommand{\Mpl}{M_{\mathrm{Pl}}}

\newcommand{\lhf}{\lambda_{h\phi}}
\newcommand{\sighf}{\sigma_{h\phi}}

\newcommand{\mk}[1]{#1}

\title{Postinflationary  vacuum instability  and Higgs--inflaton couplings     }

\author[a]{Kari Enqvist,}
\author[b]{Mindaugas Kar\v{c}iauskas,}
\author[a]{Oleg Lebedev,} % 
\author[a]{\\ Stanislav Rusak,} %
\author[a]{and Marco Zatta  } %
\affiliation[a]{University of Helsinki and Helsinki Institute of Physics, P.O. Box %
64, FI-00014, Helsinki, Finland } %
\affiliation[b]{Department of Physics, University of Jyvaskyla, P.O.
Box 35 (YFL), FI-40014 University of Jyv\"{a}skyl\"{a}, Finland}
\emailAdd{kari.enqvist@helsinki.fi} %
\emailAdd{mindaugas.m.karciauskas@jyu.fi} %
\emailAdd{oleg.lebedev@helsinki.fi} %
\emailAdd{stanislav.rusak@helsinki.fi} %
\emailAdd{marco.zatta@helsinki.fi} %
\arxivnumber{} %

\abstract{ The Higgs--inflaton coupling plays an important role in the Higgs field dynamics in the early Universe. Even a tiny coupling generated at loop level can have a dramatic effect on the fate of the electroweak vacuum. Such Higgs--inflaton interaction is present both 
at the trilinear and quartic levels in realistic reheating models.
In this work, we examine the Higgs dynamics during the preheating epoch, focusing on the effects of the parametric and tachyonic resonances. We use  lattice simulations and other numerical tools 
in our studies.  
We  find that the resonances  can induce large fluctuations of the Higgs field which  destabilize the electroweak vacuum. 
Our considerations thus provide an upper bound on quartic and trilinear interactions between the Higgs and the inflaton. We conclude that 
there exists a favorable range of the couplings within which the Higgs
field is stabilized during both inflation and preheating epochs. }

\begin{document}

\maketitle

\section{Introduction }

The discovery of the Higgs boson at the LHC~\cite{Chatrchyan:2012xdj,Aad:2012tfa} in July 2012 furnished the final piece of the Standard Model (SM) of particle physics; however, it has also raised
important new questions. One of these relates to the issue of   the electroweak vacuum stability and the fate of the Higgs field in the early
Universe, particularly during the inflationary and reheating eras 
\cite{Espinosa:2007qp}.

For the currently preferred values of the top quark mass and the strong coupling, the self-coupling of the Higgs field turns negative at a high energy scale of order 
$\mu_c \sim 10^{10}$ GeV~\cite{Buttazzo:2013uya,Bezrukov:2012sa,Alekhin:2012py} (see \cite{Espinosa:2016nld} for its gauge 
(in)dependence).  
 This would suggest that there exists
another, deeper vacuum state than the one  we currently occupy. One
finds then that the electroweak vacuum is metastable with the lifetime longer than the age of the Universe.
Although this does not pose an immediate problem, the existence of the 
deeper vacuum  raises cosmological questions. In particular,
one must explain how the Universe ended up in an energetically disfavored state and why it stayed there during inflation  \cite{Lebedev:2012sy}. Even if one fine--tunes the Higgs field initial conditions before inflation,
light scalar fields experience large fluctuations of order the Hubble rate  $H$ during the exponential expansion epoch \cite{Linde:2005ht}. Unless $H$ is sufficiently small, the Universe is overwhelmingly likely to end up in the catastrophic vacuum \cite{Espinosa:2015qea}.

These problems can be solved by coupling the Higgs field to the scalar curvature \cite{Espinosa:2007qp} or by taking into account the 
Higgs--inflaton coupling \cite{Lebedev:2012sy}. We focus on the latter possibility in this paper and neglect the effect of the non-minimal coupling to gravity.\footnote{ The effect of this term is small  close to the conformal limit.} As shown in \cite{Gross:2015bea}, Higgs--inflaton interaction is inevitable in realistic models of reheating.
Indeed, the inflaton energy must be transferred to the Standard Model 
fields which leads to a (perhaps indirect) coupling between the inflaton and the SM particles. The latter induces Higgs--inflaton interaction at loop level,
\begin{equation}
V_{H\phi} = {\lambda_{h \phi}\over 2} H^\dagger H \phi^2 
+ {\sigma_{h\phi}} H^\dagger H \phi \;,\label{VH}
\end{equation}  
where $H$ is the Higgs doublet and $\phi$ is a (real) inflaton.
Here $ \lambda_{h \phi}$ and ${\sigma_{h\phi}}$ typically receive log-divergent loop contributions and thus  require renormalization. In other words, these couplings are generated by the renormalization group (RG) evolution \cite{Gross:2015bea}. Their magnitude can be large enough to alter the Higgs 
evolution completely, in particular, by inducing a large effective Higgs mass which drives the Higgs field to zero. This mechanism is 
operative in the range 
\begin{equation}
10^{-10} < \lambda_{h \phi} < 10^{-6} \;,
\label{range}
\end{equation}
with the upper bound coming from the requirement that the Higgs--inflaton interaction preserve flatness of the inflaton potential,
and  the lower limit   dictated by the condition that the Higgs effective mass be greater than the Hubble rate during inflation.
The trilinear interaction should be subdominant, $\lhf \phi^2 \gg
\sigma_{h\phi} \phi$ so that the effective mass term does not depend 
on the sign of the inflaton field. This is usually the case in explicit reheating models  \cite{Gross:2015bea}. 

In this work, we study the effect of the above couplings after inflation. Although the Higgs--inflaton interaction  can stabilize 
the Higgs potential during inflation, during preheating its
effect can  instead be destabilizing (see also \cite{Herranen:2015ima}).  The parametric resonance~\cite{Kofman:1994rk,Kofman:1997yn} 
due to the quartic interaction $h^2 \phi^2$ and the tachyonic 
resonance~\cite{Felder:2000hj,Dufaux:2006ee}
due to the  $h^2 \phi$ term can lead to very efficient Higgs   
production. This causes     large fluctuations and the Higgs variance
$\langle h^2 \rangle$ that can exceed the critical value beyond 
which the system becomes unstable. We  find that these considerations
place important upper bounds on both $\lhf $ and $\sigma_{h\phi}$
such that the range of favored couplings (\ref{range}) reduces.

The field of Higgs dynamics in the early Universe has been very active 
in the recent years. Higgs field fluctuations during inflation 
in the metastable Universe
have been studied in \cite{Shkerin:2015exa,Kobakhidze:2013tn} and
\cite{Hook:2014uia,Kearney:2015vba,East:2016anr}. 
The Higgs condensate dynamics assuming stability of the Higgs vacuum 
were analyzed in detail  in \cite{Enqvist:2014bua,Enqvist:2015sua}.
These considerations are affected by the presence of further Higgs 
interactions which are usually not included in the Standard Model.
 The effect of the non-minimal coupling Higgs to gravity on the Higgs
dynamics was recently refined in  \cite{Herranen:2014cua}.
 In this framework, it was also noted that the resonances during preheating can destabilize the electroweak vacuum \cite{Herranen:2015ima}. The effect of the quartic Higgs--inflaton coupling on the Higgs production during preheating was considered in detail in 
 \cite{Ema:2016kpf} (see also \cite{Kohri:2016wof,Kohri:2016qqv}).
Our present work goes beyond these previous studies in that we consider 
a more realistic case of both quartic and trilinear interactions present,
which brings in new and important qualitative features. We also refine
the earlier  analysis of the pure quartic case. Finally, we discuss
implications of our findings for realistic reheating models.

This paper is organized as follows. In the next section, we present our setup. In section~\ref{sec:parametric}, we consider
the effect of the quartic Higgs--inflaton interaction on Higgs production during preheating. 
Section~\ref{sec:mixed} is devoted to the more realistic case
of both trilinear and quartic interactions present.

\section{\label{sec:setup}Framework}

In this section, we present our inflationary setup. 
For concreteness, we study the Higgs production within the  simple $m^2\phi^2$ chaotic inflation model with
\begin{equation}
m=1.3\times10^{-6}\Mpl,~~~\Mpl=1.22\times10^{19}~\text{GeV} \;,\label{m}
\end{equation}
while our results easily generalize to other large field models.
In the unitary gauge $H=(0, h/{\sqrt{2}})^{\rm T}$, the relevant Lagrangian is given by 
\begin{equation}
 \mathcal{L} = \frac{1}{2}\partial_\mu\phi\partial^\mu\phi - \frac{1}{2}m^2\phi^2 + \frac{1}{2}\partial_\mu h\partial^\mu h - \frac{\lambda_h(h)}{4}h^4 - \frac{\lhf}{4}\phi^2h^2 - \frac{\sighf}{2}\phi h^2 \;,
\end{equation}
where  the self-coupling $\lambda_h(h)$ is determined by the RG equations of the Standard Model.

 During inflation, $\phi$ undergoes a slow--roll evolution. 
  On the 
 other hand,  for $\lhf > 10^{-10}$ and a sufficiently large initial inflaton value $\phi_0 \gg M_{\rm pl}$, the Higgs mass   is dominated 
 by the inflaton interaction,  $m^{\rm eff}_h \simeq \sqrt{\lhf/2} \vert\phi\vert$
 \cite{Lebedev:2012sy}.\footnote{Here the effect of the trilinear term is negligible since we assume $\lhf \phi^2 \gg
|\sigma_{h\phi}| \phi$ during inflation.}  Then Higgs field evolves exponentially quickly to zero.
 
Not long after the end of inflation, the inflaton field undergoes oscillations
\begin{equation}
\phi(t) = \Phi(t) \cos mt \;.\label{Phicos}
\end{equation}
As long as the energy density of the Universe is dominated by the inflaton oscillations, the scale factor behaves as $a =(t/t_0)^{2/3}$ and the amplitude of oscillations decays as
\begin{equation}
\Phi(a) =\Phi_{0}a^{-3/2}.\label{Fa}
\end{equation}
For concreteness, we assume that $\phi(t)$ in eq.~(\ref{Phicos}) becomes a good approximation to the evolution of the inflaton at
\begin{equation}
\Phi_{0}\simeq0.2\Mpl.\label{F0}
\end{equation}
Soon thereafter we can accurately approximate the time dependence of $\Phi(t)$ as
\begin{equation}
\Phi(t)\simeq(3\pi)^{-1/2}\frac{\Mpl}{mt}.\label{Ft}
\end{equation}

%with $\Phi(t) \simeq \Phi_0 t_0/t$. As long as the energy density of the Universe is dominated by the inflaton oscillations, the scale factor behaves as $a =(t/t_0)^{2/3}$.  The inflaton induced Higgs mass term also oscillates which can lead to efficient Higgs production.

The inflaton induced Higgs mass term also oscillates which can lead to efficient Higgs production. As one can see in eq.~(\ref{VH}) the second term grows with respect to the first one as $\Phi(t)$ decreases due to the expansion of the Universe. Therefore the effect of the trilinear term becomes 
important at some stage even though it was negligible during inflation. 
Since the consequent effective Higgs mass term can have either sign
$\propto \sigma_{h\phi} \phi$, the tachyonic resonance becomes
effective. Both of the resonances play an important role and will
be studied in the next sections.

Before we proceed, let us clarify our assumption about the 
running coupling $\lambda_h (\mu)$,
 where $\mu$ is the renormalization scale. During the resonances, the Higgs quanta are produced coherently with the corresponding occupation numbers being very large.
Thus we may treat $h$ semi--classically. In this regime,
we may take
\begin{equation} 
 \lambda_h (\mu)= \lambda_h \left(\sqrt{ \langle h^2 \rangle}\right) \;,
\end{equation}
where $ \sqrt{ \langle h^2 \rangle}$ plays the role of the relevant 
energy scale at which the coupling should be evaluated. Since we are only
interested in the high energy regime,
in our numerical analysis we use the step--function approximation
\begin{equation} 
 \lambda_h (\mu)= 0.01 \times {\rm sign} \left(h_c^{\rm SM}- \sqrt{ \langle h^2 \rangle} \right) \;,
\end{equation}
where $h_c^{\rm SM} \sim 10^{10}$ GeV is the critical scale of the 
Standard Model at which $\lambda_h$ flips sign.

\section{\label{sec:parametric}Pure Parametric Resonance}

Let us first consider the case where the trilinear interaction is 
negligible, $\sigma_{h\phi} \approx 0$. The Higgs--inflaton interaction is quartic so that we recover the well--known 
 parametric resonance setting \cite{Kofman:1997yn}.
 
 The equations of motion for the Higgs field are quadratic in $h$ apart
from the quartic self--interaction. During the parametric resonance regime,  the effect of the latter can be approximated as $h^4 \rightarrow 6 h^2 \langle h^2 \rangle$, which is known as the Hartree approximation. In that case, the equations of motion for different momentum modes decouple. In terms of 
the rescaled Higgs momentum modes $X_k \equiv a^{3/2}h_k$, where $a$ is the scale factor, one has \cite{Kofman:1997yn}
\begin{equation}
    \ddot X_k + \omega_k^2X_k = 0 \qquad \text{with} \qquad \omega_k^2 = \frac{k^2}{a^2} + \frac{\lhf}{2}\Phi^2\cos^2(mt) + 3\lambda_h a^{-3}\langle X^2\rangle + \left(\frac{3}{2}\right)^2wH^2 \;.\label{wk-qr}
\end{equation}
 In the last term, $w = p/\rho = -\left(1 + \frac{2}{3}\frac{\dot H}{H^2}\right)$ is the equation of state parameter
of the Universe, which vanishes in the matter-like background. 
We thus neglect this term.

If the Higgs--inflaton coupling $\lambda_{h\phi}$ is substantial,
the Higgs modes experience amplification due to broad parametric resonance. The parameter characterizing the strength of the resonance is \mk{}
\begin{equation}
q(t)= \frac{\lambda_h\Phi^2(t)}{2m^2} \label{q-def}
\end{equation}
such that $q \gg 1$ corresponds to the broad resonance regime.
In this case, the modes grow exponentially leading to a large Higgs field variance $\langle h^2 \rangle$. The fluctuations can be so significant 
that they exceed the size of the barrier separating the electroweak vacuum from the catastrophic one at large field values. In this 
case, vacuum destabilization occurs. In what follows, we will estimate the corresponding critical size of $\lambda_{h\phi}$.

As was shown in~\cite{Kofman:1997yn}, in the broad resonance regime the Higgs modes evolve adiabatically away from the inflaton zero-crossings and can be described by the WKB approximation
\begin{equation}
X_k \simeq \frac{\alpha_k}{\sqrt{2\omega_k}}e^{-i\int\omega_k\ud t} + \frac{\beta_k}{\sqrt{2\omega_k}}e^{i\int\omega_k\ud t} \;,
\end{equation}
where $\alpha_k, \beta_k$ are some constants.
Adiabaticity is broken for  certain modes near the
inflaton zero-crossing, where the frequency $\omega_k$ evolves
very quickly. There 
 the system can be treated in analogy to the Schr\"odinger equation as a scattering of plane wave solutions. The adiabatic constants $\alpha_k$ and $\beta_k$
can be thought of as Bogolyubov coefficients. We assume a vacuum initial condition for the Higgs modes with $\alpha_k = 1$ and $\beta_k = 0$. The occupation number of Higgs quanta
after $j \simeq mt/\pi$ zero crossings is then
\begin{equation} 
n_k^{j+1} = \vert\beta_k^{j+1}\vert^2
\end{equation}
and can be written in terms of the corresponding  Floquet index
$\mu_k^j$ as \cite{Kofman:1997yn} 
\begin{equation}
n_k^{j+1} \simeq e^{2\pi\mu_k^j}~ n_k^{j} \;.
\end{equation}
$\mu_k^j$  can be calculated via scattering of plane waves in a parabolic potential  \cite{Kofman:1997yn},
\begin{equation}
 \mu_k^j = \frac{1}{2\pi}\ln\left[1+2e^{-\pi\kappa_j^2} + 2\sin\theta_{\mathrm{tot}}^j\sqrt{e^{-\pi\kappa_j^2}(1+e^{-\pi\kappa_j^2})}\right] \qquad \text{with} \qquad  \kappa_j^2 \equiv \frac{k^2}{\sqrt{q_j}a_j^2m^2} .\label{eq:Floquet}
\end{equation}
Here $a_j$ is the scale factor after $j$ zero crossings.
The term $\sin\theta_{\mathrm{tot}}$ is determined from the phase accrued by the modes and behaves in a stochastic manner for different momenta (see~\cite{Kofman:1997yn}).
We take it to be zero for our estimates and use the consequent
average value of the Floquet index. 
The occupation numbers at late times ($a_j \gg 1$) can  then be approximated by
\begin{equation}
 n_k^{j+1} \simeq \frac{3^j}{2}e^{-\bar\mu_j\frac{k^2}{m^2}} \qquad \text{where} \qquad \bar\mu_j = \frac{\Mpl}{\sqrt{3 \pi q_0}\Phi_0}\; a_j \;.
\end{equation}
Here the factor of $1/2$ is due to the vacuum fluctuations in the initial state, although it plays no tangible role in our analysis.

Using the saddle point approximation, 
the Higgs field variance can be written as
\begin{equation}
 \langle h^2\rangle \simeq \int \frac{\ud^3 k }{(2\pi a)^3}\frac{n_k}{\omega_k} \simeq \frac{3^jm^2 \kappa_\mathrm{max}^3}{2^{5/2}e\pi^{3/2}a^3 \sqrt{q}} \;,
 \label{h2}
\end{equation}
where $\kappa_\mathrm{max}^2 = \bar\mu_j^{-1}$ is the momentum in   units of $m$ which contributes most significantly. Here 
we have assumed that $\omega_k$ is dominated by the 
inflaton--induced term.\footnote{This assumption does not have a significant numerical impact on our main results. } 
Already after the first zero-crossing $ \langle h^2\rangle$ exceeds the critical scale $\sim 10^{10}$ GeV of the Standard Model and therefore the Higgs self-coupling  $\lambda_h = \lambda_h\left(\sqrt{\langle h^2\rangle}\right)$
can be taken to be negative from the beginning. For our analysis, we take $\lambda_h = -10^{-2}$ at large field values. Note that the fact
that $ \langle h^2\rangle$ exceeds the SM critical scale does not necessarily lead to vacuum destabilization since the presence of the Higgs--inflaton coupling pushes the barrier separating the two vacua to larger values of order 
\begin{equation}
h_c \sim \sqrt{\lambda_{h\phi}\over \vert\lambda_h\vert} \;\vert\phi\vert \;. 
\label{barrier}
\end{equation}
However, the position of the barrier is modulated by $|\cos mt|$ so it is not immediately clear what vacuum stability would require.

To derive the stability condition, one can use the following reasoning.
Around each inflaton zero crossing, the effective Higgs mass squared is 
dominated by the Higgs self-interaction term $\lambda_h \langle h^2\rangle$. Such a tachyonic term leads to  exponential amplification of the Higgs field by a factor of order $e^{m_h^{\rm eff} \Delta t}$,
where $m_h^{\rm eff} $ is the modulus of the effective Higgs mass term and $\Delta t$ is the  (short) period during which the Higgs self-interaction term dominates. $\Delta t$  is given explicitly by
 \begin{equation}
 |\Delta t| < \sqrt{\frac{6|\lambda_h|\langle h^2\rangle}{\lhf\Phi^2 m^2}} \;. 
\end{equation}
The tachyonic amplification is insignificant as long as $m_h^{\rm eff} \Delta t$ does not exceed unity, that is,
\begin{equation}
\sqrt{3|\lambda_h|\langle h^2\rangle} \; |\Delta t|<1 \;.
\end{equation}
Clearly, this condition eventually gets violated since $\Delta t$ grows
as the inflaton amplitude $\Phi$ decreases. However, if the resonance ends before this takes place, no destabilization occurs. 
Using 
$\lhf\Phi^2 \simeq 2m^2$ at the end of the resonance
~\cite{Kofman:1997yn}, 
one finds that the stability condition can be written as
\begin{equation}
 \lhf < \frac{6\pi^3}{(\ln 3)^2}\frac{m^2}{\Mpl^2}\left[\ln\left(\frac{16e\pi^{3/2}}{9|\lambda_h|}\right) + {3\over2} \ln\left(\sqrt{\frac{\lhf}{16 \pi}}\frac{\Mpl}{m}\right)\right]^2 \ \simeq 3 \times 10^{-8}.
 \label{main-bound}
\end{equation}
Here we have neglected a smaller additive constant in the square brackets.
If this condition is violated, the Higgs field grows explosively
since the amplification factor $e^{m_h^{\rm eff} \Delta t}$ increases
with  $\langle h^2\rangle$ itself.  This leads to fast vacuum destabilization.
On the other hand, if this condition is satisfied, it implies that
the Higgs potential is dominated by the inflaton coupling term on the average and $h$ does not fluctuate beyond the barrier (\ref{barrier}).
This result is consistent with the bound obtained in \cite{Ema:2016kpf}.

 \begin{figure}
 \center
 \includegraphics[width=.9\textwidth]{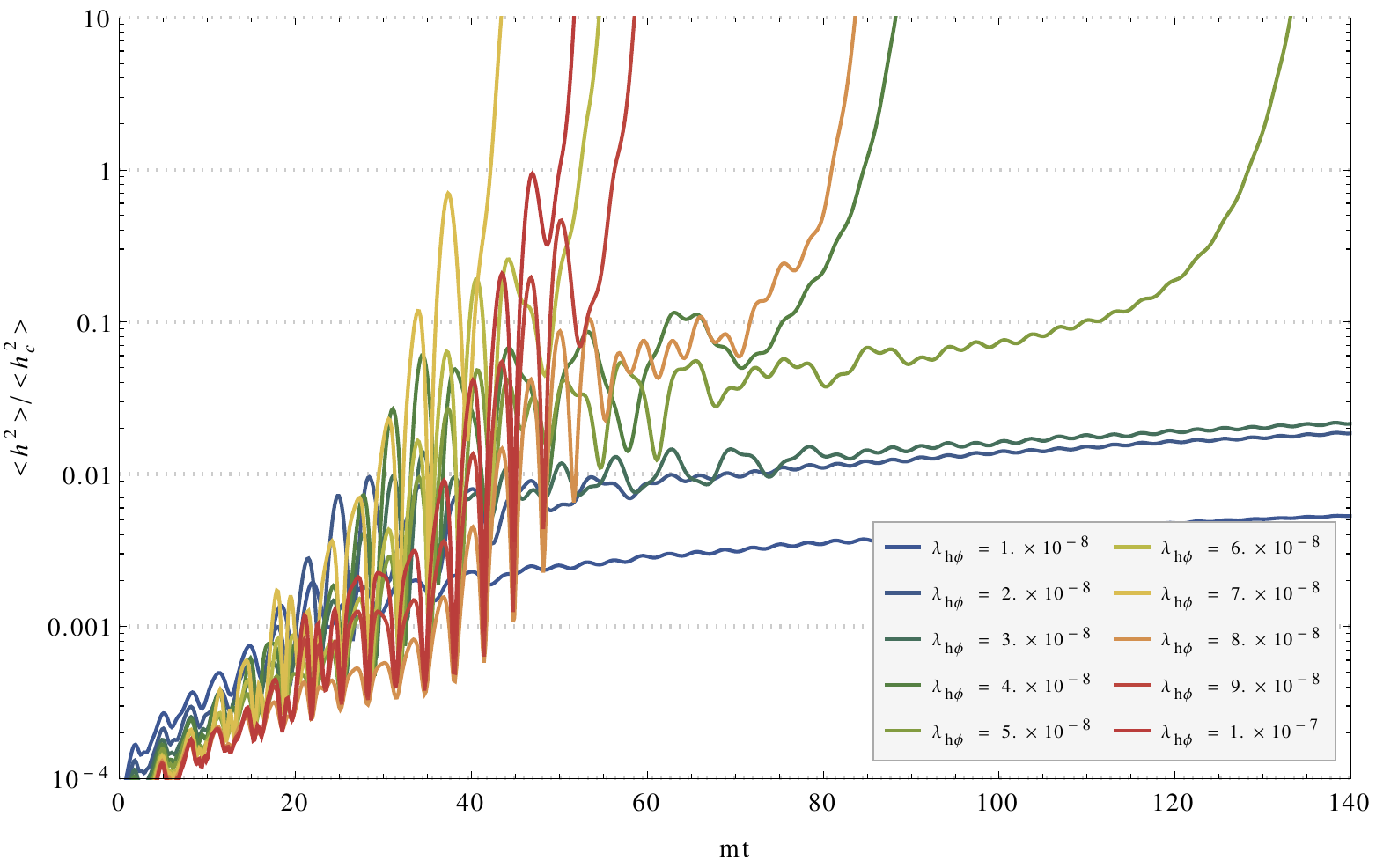}
 \caption{\label{fig:h_vs_hc}  Time evolution of the Higgs  fluctuations scaled by the location of the potential barrier $h_c = \sqrt{\frac{\lhf}{|\lambda_h|}}\Phi$, 
for different $\lhf$. The $\lambda_h h^4$ term is treated in the {\it Hartree approximation}. }
\end{figure}
 
Figure~\ref{fig:h_vs_hc} shows our numerical evolution of the Higgs fluctuations for different values of $\lhf$. To produce this plot we have solved the mode equations in the Hartree approximation using Mathematica software. We see that for $\lhf$ greater than
a few times $10^{-8}$, the Higgs field grows above the critical value and blows up at $mt> 40$. The destabilization time however should not
be taken at face value since the   Hartree approximation turns out
to be rather crude for this purpose.

\begin{figure}
 \center
 \includegraphics[width=.85\textwidth]{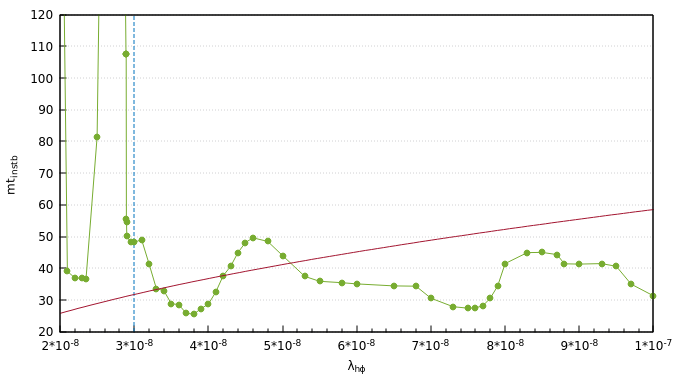}
 \caption{\label{fig:s0-destb-1} Vacuum destabilization time versus $\lhf$ (green curve) with LATTICEEASY.
Points below the red line correspond to the active parametric resonance. Our theoretical upper bound on  $\lhf$ is marked by the vertical dashed line. }
\end{figure}

We have also performed  a more sophisticated  lattice simulation which takes into account
the Higgs self--interaction without resorting to the Hartree approximation. We used the  LATTICEEASY package \cite{Felder:2000hq}
for this purpose. In our simulations, we choose  the  box size of $10/m$ (the $L$ parameter of LATTICEEASY) with $64$ grid points per edge (the $N$
parameter). We have checked that a larger and finer grid does not change the results significantly.
In Figure~\ref{fig:s0-destb-1}, we plot the destabilization time versus $\lhf$. The green curve shows   $mt$ at which the system is destabilized, that is, 
the Higgs field variance blows up. The red line marks the end of the resonance such 
that the points below it correspond to vacuum destabilization during the resonance   as
studied in this section.\footnote{The EW vacuum can be destabilized at later times 
as seen in Figure~\ref{fig:s0-destb-1}. This is however a different phenomenon which  we  consider in subsequent sections.} 
Our theoretical bound on $\lhf$ is marked by the vertical dashed line. We see that
the latter describes the general situation reasonably well and $\lhf$ above
$3 \times 10^{-8}$ typically leads to vacuum destabilization during the resonance.
On the other hand, we also see the limitations of our approach. In particular,
Figure~\ref{fig:s0-destb-1} shows 
that  the strength of the resonance does not behave monotonically with $\lhf$. 
This is expected since 
we have taken the term $\sin\theta_\mathrm{tot}$ to be zero, whereas in reality it either enhances or suppresses the resonance
 such that there can be certain values of $\lhf$ satisfying our bound yet leading to an unstable configuration.
Formally, the area around $\lhf \sim 2 \times 10^{-8} $ appears to be stable during the resonance
and the destabilization occurs shortly after the resonance. However, 
one can classify this region as unstable since in reality the end of the resonance is not sharply defined due to various approximations we have made.  Apart from these complications, we find that our simple considerations
give a fairly good description of the system behavior during the parametric resonance.

Comparing Fig.~\ref{fig:h_vs_hc} and Fig.~\ref{fig:s0-destb-1},
one finds that the commonly used Hartree approximation overestimates
the destabilization time. This is to be expected since the quantity
$h^4$ experiences greater fluctuations than  $h^2 \langle h^2 \rangle$
does. Nevertheless certain questions  such as the effect 
of perturbative Higgs are easier addressed using
our Mathematica routine which employs the Hartree approximation.
Hence we use both numerical approaches.

Figure~\ref{fig:s0-destb-1} also shows that the late time behaviour (beyond the resonance)
 of the Higgs fluctuations is important, which we discuss in section \ref{late}.

\begin{figure}
 \includegraphics[width=.5\textwidth]{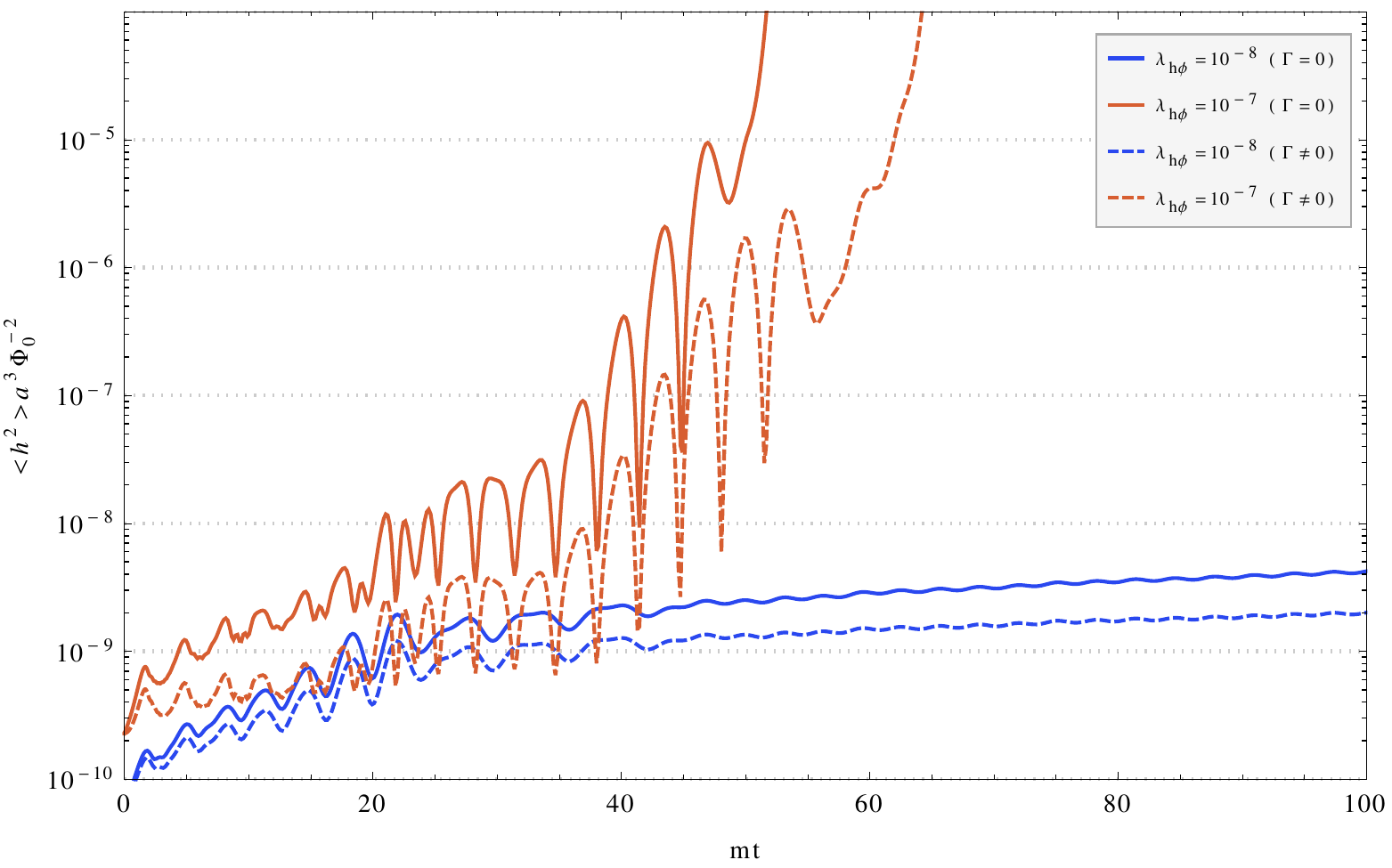}
 \includegraphics[width=.5\textwidth]{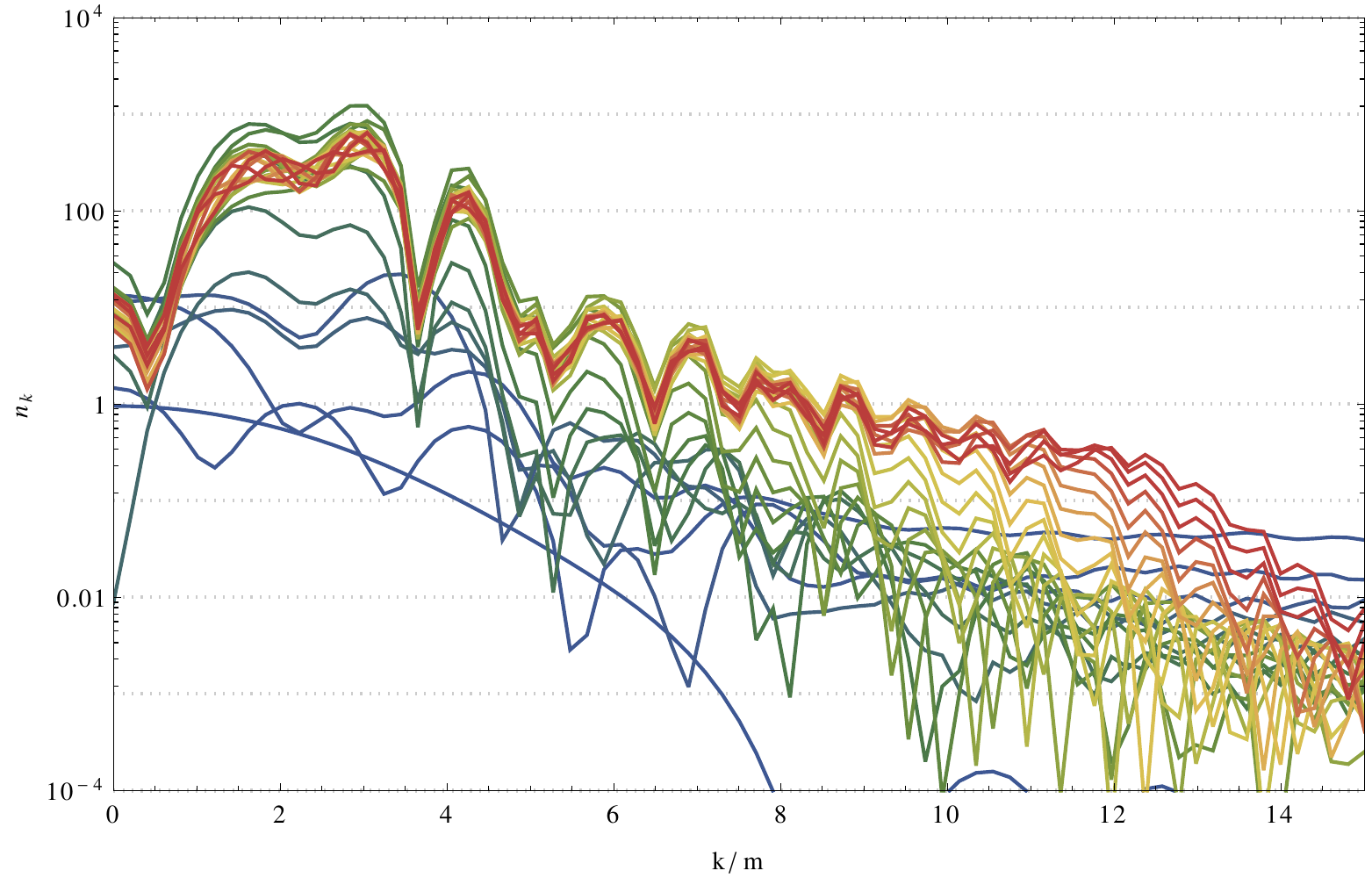}
 \caption{\label{fig:nk}   Left: effect of the perturbative Higgs 
 decay $h\rightarrow t \bar t$. Right: example of the occupation number evolution for different momenta with $\lhf = 3\times 10^{-8}$. Blue (red) curves correspond to early (late) times. (The {\it Hartree approximation} is employed). }
\end{figure}

So far our discussion has ignored perturbative decay of the Higgs quanta, which reduces the efficiency of the resonance and can potentially invalidate our conclusions.
 The main decay channel is provided by the top quarks which are
effectively massless for our purposes. The corresponding decay width is
\begin{equation}
\Gamma (h\rightarrow t \bar t)= {3 y_t^2 m_h^{\rm eff} \over 16 \pi} \;.
\end{equation}
Taking  $y_t( m_h^{\rm eff}  ) \sim 1/2$,  $m_h^{\rm eff}\simeq \sqrt{\lhf /2} \; \vert \phi \vert $ 
and averaging $\vert \cos mt \vert$, we find that the perturbative decay    reduces the number of the Higgs quanta by a factor  2 or so in the region
of interest ($\lambda_{h\phi} \sim 10^{-8}$), see \mk{the left panel of} Figure~\ref{fig:nk}. Therefore it does not
significantly affect our bound on $\lambda_{h\phi}$.
On the other hand, for larger $\lhf \sim 10^{-7}$, the Higgs decay can reduce $\langle h^2 \rangle $ by an order of magnitude thus delaying
(but not avoiding) vacuum destabilization. 
 
For completeness, in the right panel of Figure~\ref{fig:nk}, we present a typical example of the occupation number evolution for different momenta. We find
that at late times the Higgs field is typically dominated by the modes with momenta $k \sim m$.

\section{\label{sec:mixed}Effect of  the Trilinear Interaction}

The  trilinear Higgs--inflaton interaction brings in an additional effective mass term whose sign
oscillates in time. This results in the  tachyonic resonance \cite{Felder:2000hj} which
amplifies the Higgs fluctuations. We find that the effect is important and cannot be neglected.

The parametric and tachyonic resonances have been studied separately in detail. In realistic models,
both of them are present at the same time, yet their combined effect is not well understood (see however \cite{Dufaux:2006ee},\cite{Lachapelle(2009)preh})\footnote{We also note that the range of parameters considered in these papers  is very different from that of interest here.}. In particular, the Higgs field goes
through a sequence of exponential  amplification periods and  plateaus.
In what follows,  we  study some of the important aspects of the system and obtain the
corresponding bound on $\sigma_{h\phi}$.

\subsection{Equations of Motion}

 The trilinear
interaction introduces an additional oscillating contribution to the
effective Higgs mass. The frequency of this contribution is half the
frequency of the quartic interaction. In particular, the Higgs dispersion
relation in eq.~(\ref{wk-qr}) becomes
\begin{equation}
\omega_{k}^{2}=\left(\frac{k}{a}\right)^{2}+\sigma_{h\phi}\Phi\left(t\right)\cos  mt +\frac{1}{2}\lambda_{h\phi}\Phi^{2}\left(t\right)\cos^{2}  mt  +3\lambda_{h}a^{-3}\left\langle X^{2}\right\rangle ,\label{mx-EoM}
\end{equation}
where, as in eq.~(\ref{wk-qr}), we have neglected
terms proportional to $\dot{H}\sim H^{2}$. These terms become small,
as compared to $m^{2}$, soon after the end of inflation. Let us  introduce  
\begin{eqnarray}
p\left(t\right) & \equiv & 2\frac{\sigma_{h\phi}\Phi\left(t\right)}{m^{2}}\;,\label{p-def}\\
\delta m^{2}\left(t\right) & \equiv & 3\lambda_{h}a^{-3}\left\langle X^{2}\right\rangle \;.\label{dm2-def}
\end{eqnarray}
\mk{and the $q(t)$ parameter, which is defined in eq.~(\ref{q-def})}. Then the equation of motion for the (rescaled) Higgs field can be
written as
\begin{eqnarray}
\frac{\mathrm{d}^{2}X_{k}}{\mathrm{d}z^{2}}+\left[A\left(k,z\right)+2p\left(z\right)\cos 2z +2q\left(z\right)\cos 4z+\frac{\delta m^{2}\left(z\right)}{m^{2}}\right]X_{k} & = & 0,\label{mx}
\end{eqnarray}
where
\begin{align}
z & \equiv\frac{1}{2}mt \;,\\
A\left(k,z\right) & \equiv\left(2\frac{k}{am}\right)^{2}+2q\left(z\right)\;. \label{A-def}
\end{align}
This differential equation reduces to the Whittaker--Hill equation
if the Universe expansion and \mk{the} Higgs self--interaction are neglected.
Its solutions exhibit the resonant behavior similar to those of the Mathieu equation, although the situation is more complicated due to
the presence of two parameters $p$ and $q$.
According to the Floquet theorem, a general solution of the Whittaker--Hill equation
can be written as
\begin{equation}
X\left(z\right)=\rho_{1}\mathrm{e}^{\mu z}y\left(z\right)+\rho_{2}\mathrm{e}^{-\mu z}y\left(-z\right),\label{WH-gensol}
\end{equation}
where $\rho_{1}$ and $\rho_{2}$ are integration constants, $y\left(z\right)$
are periodic functions of period $\pi$ and $\mu$ is a characteristic
exponent, or Floquet exponent, which in general is a complex number. 
When $\mu$ attains a real part, the solution grows exponentially.
We discuss the most important properties of these solutions
in Appendix~\ref{sec:WH-Eqn}. In particular, the stability chart of
the  Whittaker--Hill equation is quite  different from that 
of the Mathieu equation in the parameter range of interest (see 
fig.~\ref{fig:WH-stb-istb}).

In reality,  the Universe expansion cannot be neglected and leads to the end of the resonance.
Hence the Whittaker--Hill equation only provides a simple
approximation to the equations of motion.
 The duration of the resonance is essential for our considerations since it determines the size of $\langle h^2 \rangle$. Let us consider 
 it in detail.

\subsection{Duration of the Resonance\label{subsec:mx-nolh}}

%Let us first neglect the Universe expansion and Higgs self--interaction.
%In that case,
% parameters $A$,
%$p$, $q$ are constant and  eq.~(\ref{mx}) reduces
%to the Whittaker-Hill equation. The solutions of this equation and
%their stability are discussed in Appendix~\ref{sec:WH-Eqn}. 
%The behavior of these solutions can be understood in terms
%of the stability charts in  figure~\ref{fig:WH-stb-istb} that show 
%which sets of parameters $A$, $p$ and $q$ lead to exponentially growing
%solutions. Even though these parameters are not constant in reality, such charts are useful and help estimate the time when the
%resonance terminates.
\mk{}

The essential difference between the solutions of eq.~(\ref{mx})
in an expanding and static Universes is that in the former case
the boundaries between the stability and instability regions are no longer
clearly defined: they are smeared \cite{Kofman:1997yn}. Despite
this fact, we will use  figure~\ref{fig:WH-stb-istb}
as a helpful illustration. For that matter, the time dependence
of $A\left(k,z\right)$, $p\left(z\right)$ and $q\left(z\right)$
can be introduced adiabatically: as they evolve, one can think of them
as tracing a trajectory in the three dimensional space, crossing through
stable and unstable regions. Once these parameters decrease substantially
and the trajectory converges  to the lowest stable region, the resonance
ceases.

\begin{figure}

\begin{centering}
\includegraphics[scale=0.5]{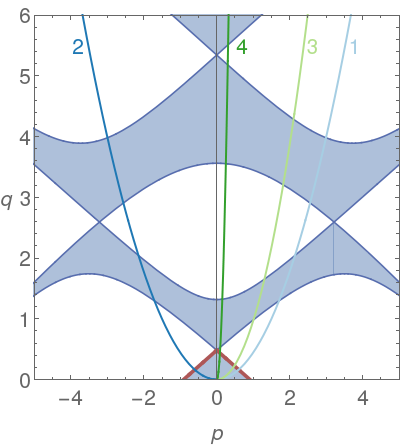}
\par\end{centering}
\caption{\label{fig:stb-instb-k0}Stability (shaded) and instability
(white) regions of the Whittaker-Hill equation for  $A=2q$. 
The labeled curves describe   evolution of $p(t),q(t)$
for  different   values of $\sigma_{h\phi}$
and $\lambda_{h\phi}$: 
(1) $\sigma_{h\phi}= 8\times 10^{-11} M_{\rm Pl},\; \lambda_{h\phi} = 1.5 \times 10^{-8}$; 
(2) $\sigma_{h\phi}= -8\times 10^{-11} M_{\rm Pl},\; \lambda_{h\phi} = 1.5 \times 10^{-8}$; 
(3) $\sigma_{h\phi}= 7\times 10^{-11} M_{\rm Pl}, \;\lambda_{h\phi} = 2.5 \times 10^{-8}$; 
(4) $\sigma_{h\phi}= 1\times 10^{-11} M_{\rm Pl}, \;\lambda_{h\phi} = 3 \times 10^{-8}$.
The boundary of the last stability region  around $p=0$ is marked in red.}
\end{figure}

The definition of the parameter $A$ in eq.~(\ref{A-def}) contains
two terms, which are time dependent. To compute the duration of the
resonance, we first estimate the relative size of these two contributions
at the end of the resonance. Using eq.~(\ref{Fa}) and the definition
of $q$ in eq.~(\ref{q-def}), we can write
\begin{equation}
\frac{\left(2k/a_{f}m\right)^{2}}{2q_{f}}\sim\frac{\left(m/\Phi_{0}\right)^{4/3}\left(k/m\right)^{2}}{\lambda_{h\phi}^{2/3}q_{f}^{1/3}}~,
\end{equation}
where the subscript $f$ refers to  values at the end of the resonance.
Typically, the excited modes  towards the end of the resonance are $k/m\sim1$.
 Since $ m/\Phi_{0} \sim 5 \times 10^{-6}$,
we have 
\begin{equation}
\frac{\left(2k/a_{f}m\right)^{2}}{2q_{f}}\simeq\frac{10^{-7}}{\lambda_{h\phi}^{2/3}q_{f}^{1/3}}~.\label{kq}
\end{equation}
\mk{In this work we are interested in the range of values of $\lambda_{h\phi}$ given in equations (\ref{range}) and (\ref{main-bound}). Taking also $q_{f}^{1/3}\sim 1$, the above ratio lies in the range $10^{-2}...1$.}
%In this section, we are interested in the range of $\lambda_{h\phi}$
%which leads to a stable Higgs configuration without the trilinear term.
%, that is 
%$\lambda_{h\phi}\sim10^{-10}-10^{-8}$. 
%Taking $q_{f}^{1/3}\sim 1$, 
%we find that the above ratio lies in the range $10^{-2}...1$.
 That is, at
the end of the resonance,
the  $q$--term dominates and it suffices for our purposes to consider 
the evolution of the $k=0$ mode only. This restricts our parameter space to the plane $A=2q$.
The stability and instability regions for constant
 $p$,$q$ can be obtained  by the methods discussed
in  Appendix~\ref{sec:WH-Eqn}. The result is shown in figure~\ref{fig:stb-instb-k0}, where the labeled curves display the trajectories 
$p(t),q(t)$ for different $\lambda_{h\phi}$ and $\sigma_{h\phi}$.
The vertical line $p=0$  corresponds to the parametric resonance and
one recovers the
standard results of ref.~\cite{Kofman:1997yn}. 

\begin{figure}
\begin{centering}
\includegraphics[scale=0.5]{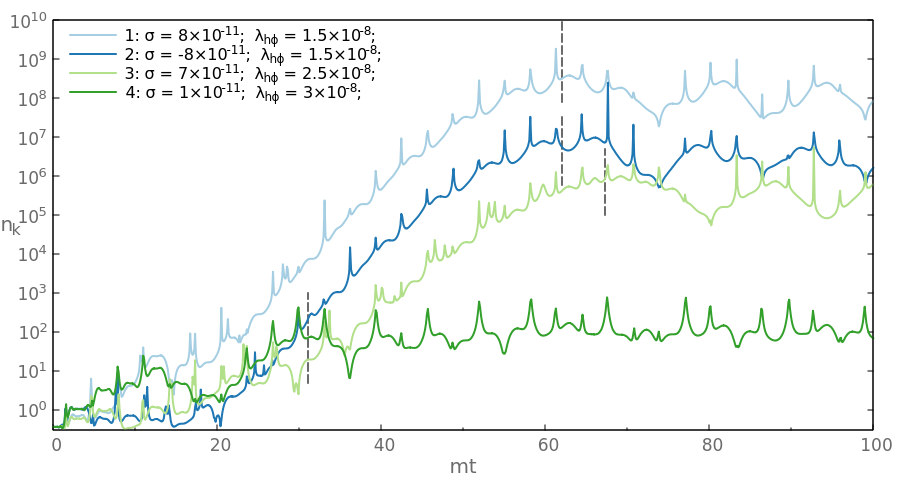}
\par\end{centering}
\caption{\label{fig:stb-instb-k}Evolution of the occupation numbers for  the mode $k/m=0.63$ with LATTICEEASY. Different color lines  correspond to models  with different values
of $\sigma$ and $\lambda_{h\phi}$ (same as in fig.~\ref{fig:stb-instb-k0}).   The vertical
dashed lines show the end of the resonance according to eq.~(\ref{mt34f}),
apart from model 4 for which the standard result $q_{f}=1$ holds 
\cite{Kofman:1997yn}. The Higgs self--interaction is set to zero, $\lambda_h=0$.}
\end{figure}

\begin{figure}
\begin{centering}
\includegraphics[scale=0.5]{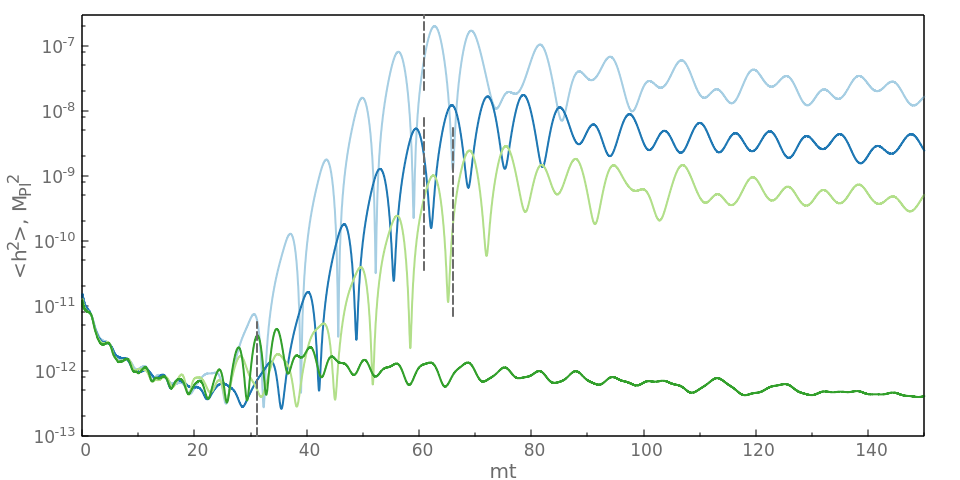}
\par\end{centering}
\caption{\label{fig:stb-instb-h} Evolution of $\langle h^{2}\rangle $ for the models of figure~\ref{fig:stb-instb-k} and  $\lambda_h=0$.
 Due to the Universe expansion, $\langle h^{2}\rangle $ decreases 
 when the resonance is not active.}
\end{figure}

The resonance stops when   $q\left(z\right)$ and $p\left(z\right)$
reach the last stable region   around $p=0$ and $q=0$
in figure~\ref{fig:stb-instb-k0}. To estimate the time when this
happens, we approximate the boundary of the lowest stable region by
a linear relation $q=0.48-0.53\left|p\right|$. This approximation
is shown by bold red lines in figure~\ref{fig:stb-instb-k0}. One
has to keep in mind however, that in the expanding universe the boundaries
between stable and unstable regions are smeared. Hence, even when a
trajectory in $\left(q,p\right)$ parameter space reaches the last
stable region, the resonance continues for some time, depending on
the phase. Thus, the end of the resonance corresponds to
\begin{equation}
q_{f}=0.48\left(1-\delta\right)-0.53\left|p_{f}\right|,\label{bndr}
\end{equation}
where $\delta$ is a ``fudge" factor to be determined from simulations.
Our results show that 
 $\delta$ varies from $0$ to about $1/4$.
An analogous result for the parametric resonance was obtained  in
 \cite{Kofman:1997yn}, in which case the resonance stops
 somewhere in the range  $1\le q_{f}\le4/3$.\footnote{Note that the definition of $q$ in ref.~\cite{Kofman:1997yn}
differs from ours by a factor of $1/4$. } 
In our parameter range, we find that $\delta\simeq0.1$ gives a good approximation for most cases.

Using eqs.~(\ref{p-def}), (\ref{q-def})
and eq.~(\ref{Ft}), we find
\begin{equation}
mt_{f}\simeq\left[3.25\frac{\left|\sigma_{h\phi}\right|/\Mpl}{\lambda_{h\phi}}\left(\sqrt{1+0.77\lambda_{h\phi}\frac{m^{2}}{\sigma_{h\phi}^{2}}}-1\right)\right]^{-1}.\label{mt34f}
\end{equation}
In figures~\ref{fig:stb-instb-k} and \ref{fig:stb-instb-h}, we plot
numerical LATTICEEASY computations of occupation numbers $n_{k}$
with $k\approx0.63m$ and $\left\langle h^{2}\right\rangle $ for
several models.\footnote{Note that, as expected,   $n_k$ starts growing when
$p(t),q(t)$ reach the relevant instability region. In particular,
for curve 2 the growth begins at $mt\sim 20$. }
 The values of $t_{f}$ from eq.~(\ref{mt34f}) are
shown by dashed vertical lines. We conclude that  the agreement is quite good.

Eq.~(\ref{mt34f}) does not apply for very small values of $\sigma_{h\phi}$ such that the tachyonic resonance is inefficient.
In particular, the amount of time the system spends in the last instability region
(just above the red line in fig.~\ref{fig:stb-instb-k0}) is so small
that no substantial amplification occurs.
For such models, the dynamics of the resonance are  close to those
of the pure parametric case \cite{Kofman:1997yn}.
It is also worth recalling that $\lambda_{h\phi}$ in eq.~(\ref{mt34f})
is not allowed to  be too small so that the ratio in eq.~(\ref{kq}) is below unity.

\subsection{Vacuum Destabilization by a Mixed Resonance }

As in the parametric resonance case, the Higgs field fluctuations
can grow large enough so that the system moves over to the catastrophic vacuum. This transition is facilitated by the presence of the trilinear term which results in very large Higgs occupation numbers.
In what follows, we study the destabilization effect due to 
$\sigma_{h\phi}$. That is, we choose $\lambda_{h\phi}$ for which 
the system is stable and analyze how large a $\sigma_{h\phi}$ one can
add without destabilizing the vacuum. As before, we focus on the 
destabilization during the resonance, i.e. before $t_{f}$   in eq.~(\ref{mt34f}).

\subsubsection{\label{sec:sbound}Simplified Bound on $\sigma_{h\phi}$}

The analysis of the mixed trilinear--quartic case is substantially more complicated than the pure quartic case. As seen from the stability chart, the system goes through a series of stable and unstable regions with a varying exponent $\mu (t)$. We will thus content ourselves with only an order of magnitude estimate of the critical $\sigma_{h\phi}$.

Towards the end of the resonance, the Higgs--dependent potential
is dominated by the trilinear  term ${1\over2} \sigma_{h\phi} \phi h^2$  since the  quartic interaction decreases faster with time. The destabilization occurs when this term becomes overtaken by the Higgs
self--interaction ${1\over4} \lambda_{h}  h^4$. Therefore, 
one can estimate the
critical variance by 
\begin{equation}
\langle h^2 \rangle_{\rm cr} \sim {2 |\sigma_{h\phi}| \Phi \over |\lambda_h |} \;,
\end{equation}
where the Hartree approximation has been  used and the oscillatory 
behavior   of $\phi$ has been ignored. 

On the other hand, the Higgs variance as a function
of time can be calculated via the occupation numbers as in (\ref{h2}).
The dominant contribution is given by modes around the comoving momentum $k_*$ which
maximizes $n_k$. For the parameter range of interest,   we find that $k_* \sim m $ towards the end
of the resonance and the width of the $k$-distribution is of
order $k_*/2$. The corresponding $n_{k_*}$ is a rather complicated function of time containing sections where it undergoes an exponential increase. For our purposes, we simply interpolate it by $e^{ \mu_* mt}$ with some effective exponent $\mu_*$. We then obtain
\begin{equation}
\langle h^2 \rangle \simeq {\Delta k_* k_*^2 \over a^3}\; {n_{k_*}\over \omega_{k_*}} \sim {m^3 \over 2 a^3 } \; {e^{ \mu_* mt} \over 
\sqrt{|\sigma_{h\phi}| \Phi} }\;.
\end{equation}

The destabilization occurs if $\langle h^2 \rangle$ reaches the critical
value during the resonance. The latter stops around $2|\sigma_{h\phi}| \Phi_{\rm end} \simeq m^2$. Taking this into account and dropping order 
one constants,  one finds 
 \begin{equation}
 |\sigma_{h\phi}| < { m^2 \over    M_{\rm pl} } \times
 {1\over \mu_*}\: \ln {a^3_{\rm end} \over |\lambda_h| } \sim  10^9 \; {\rm GeV} \;,
\end{equation}
with $a_{\rm end}=(\Phi_0/\Phi_{\rm end})^{2/3}$ being the scale factor at the end of the resonance.
Here  we take  a typical value\footnote{This is supported by our numerical analysis. }   $\mu_* \sim {\cal O}(10^{-1})$ (cf. fig.~\ref{fig:WH-stb-istb}). Note that the main
$\sigma_{h\phi}$--dependence of the result comes from the duration of the resonance, $mt_{\rm end} \sim \sigma_{h\phi} \times M_{\rm pl}/m^2$,
while that of $\mu_*$ and $\ln a_{\rm end}$ is milder.

Although this 
estimate is very crude, we find that the bound is within a factor of a few 
from our numerical results. Here we have neglected both
the $\lhf$--dependence and the dependence on the sign of 
$\sigma_{h\phi}$.  

Note that both the $\lhf$ and $\sigma_{h\phi}$ bounds do not appear 
to depend explicitly on the critical scale of the Standard Model.
This dependence is hidden in our assumption about $\lambda_h$ at the energy scales of interest. As long as $\lambda_h\sim -10^{-2}$ in that
range, our bounds apply. 

Finally, we have considered a chaotic $\phi^2$
inflation model which fixes a large $H_{\rm infl}\sim 10^{14}$ GeV.
For models with small $H_{\rm infl}< 10^{10}$ GeV, the Higgs fluctuations during inflation are not dangerous and the Higgs--inflaton coupling
can be set negligibly small.\footnote{Such models however do not solve the problem of the Higgs initial conditions at the beginning of inflation.}

\begin{figure}
\begin{centering}
\includegraphics[scale=0.6]{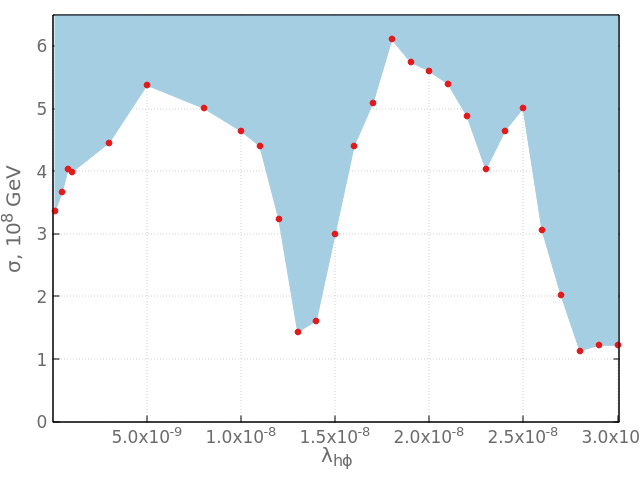}
\par\end{centering}
\caption{\label{fig:destb} Upper bound on $\sigma_{h\phi}>0$ from
LATTICEEASY simulations. 
In the shaded region, the Higgs vacuum is destabilized during the
resonance.}
\end{figure}

\begin{figure}
\begin{centering}
\includegraphics[scale=0.6]{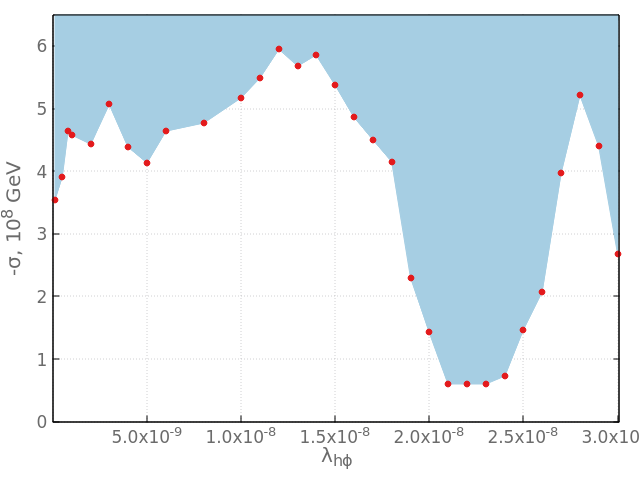}
\par\end{centering}
\caption{\label{fig:destb-neg} Upper bound on $|\sigma_{h\phi}|$
for negative $\sigma_{h\phi}$ from
LATTICEEASY simulations. 
In the shaded region, the Higgs vacuum is destabilized during the
resonance.}
\end{figure}

\subsubsection{\label{sec:sbound1}Simulation Results}

Our  LATTICEEASY  simulations show that 
the bound on $\sigma_{h\phi}$  depends both on $\lhf$ and the sign
of $\sigma_{h\phi}$. The latter is due to the fact that even though
the Whittaker-Hill equation enjoys the symmetry $z \rightarrow
z+\pi/2, \; p\rightarrow -p$, the time translation invariance is broken by the Universe expansion.
Figs.~\ref{fig:destb} and \ref{fig:destb-neg} display the   bounds on $\sigma_{h\phi}$
as a function of $\lhf$. We see that the upper bound varies between
$10^8$ GeV and $6\times 10^8$ GeV in the region of interest.

We should note that  these plots are somewhat simplified in that it is
tacitly implied that $|\sigma_{h\phi}|$ below the critical value leads
to a stable system. In practice, this is not always the case  and the 
destabilization time can be a non-monotonic function of $\sigma_{h\phi}$. However, these effects do not change our results drastically.

\section{\label{late}Comments on the Late Time Behavior}

So far we have discussed vacuum destabilization during the resonance.
The initial stage of preheating is dominated by
a single process, that is, resonant Higgs production. At later stages,
other processes such as rescattering, thermalization, etc. become important. 

As seen in fig.~\ref{fig:s0-destb-1}, the Higgs vacuum can be destabilized much after the end of the resonance. The simple reason for it
is that $\sqrt{\langle h^2 \rangle}$ and the position of the barrier $h_c \propto \Phi$ scale differently in time. If only the quartic coupling is present,
\begin{eqnarray}
 h_c &\propto& a^{-3/2} \;, \nonumber\\
 \sqrt{\langle h^2 \rangle} &\propto& a^{-\alpha} \;,
\end{eqnarray}
 where $\alpha$ is between $1$ and $3/4$, depending on which
 $k$--modes dominate $\langle h^2 \rangle$. This can be seen 
 from the first equality in (\ref{h2}) and the fact that the comoving
 occupation numbers are constant after the end of the resonance,
 while the $\omega_k$ scaling depends on the balance between $k^2/a^2$ and
 the inflaton--induced mass term.
In any case, $\sqrt{\langle h^2 \rangle} $ decreases slower in
time than $h_c$ does so that after a sufficiently long period 
the Higgs fluctuations go over the barrier. Fig.~\ref{fig:s0-destb-1}
shows that the relevant time scale is of order 100$mt$. Analogous
considerations also apply to the mixed trilinear--quartic case.

However, the true dynamics of the system on a larger time scale are complicated. The Higgs interacts with other fields of the Standard Model which becomes important after the resonance.
As noted in  \cite{Ema:2016kpf}, thermalization effects can generate a thermal mass
term for the Higgs  thereby stabilizing the vacuum. Also, non--perturbative
production of particles via the Higgs couplings can reduce
$\langle h^2 \rangle$ \cite{Enqvist:2015sua}. 
 These effects are subtle and require a careful investigation which
 is beyond the scope of our present work. On the other hand,
 the resonance regime is quite well understood and thus we believe
 our bounds on $\lhf$ and $\sigma_{h\phi}$ are solid.

\section{Implications for Reheating Models}

In this section, we consider implications of our bounds for  model
parameters of representative reheating scenarios.
We choose  two examples considered in \cite{Gross:2015bea}: reheating
via right-handed neutrinos  and reheating via non--renormalizable operators. 

In general one expects the Higgs--inflaton couplings to be present 
already at the tree level. However, if they are for some reason suppressed, $\lhf$ and $\sigma_{h\phi}$ are generated by loop corrections. Therefore, the loop--induced couplings
 can be regarded as the corresponding lower bound. In what follows, we consider two conservative scenarios in which $\lhf$ and $\sigma_{h\phi}$ are entirely due to loop effects. 
 
\subsection{Reheating via Right--handed Neutrinos}

In this model, the inflaton decays into heavy right--handed neutrinos
which subsequently decay into SM particles. This option is attractive since the inflaton--neutrino coupling is allowed already at the renormalizable level. 
 The relevant interaction terms are
\begin{equation}
-\Delta {\cal L} = {\lambda_\nu \over 2} ~\phi \nu_R \nu_R + {y_\nu } ~
\bar l_L \! \cdot \! H^* \, \nu_R + {M\over 2} ~ \nu_R \nu_R + {\rm h.c.}~, 
\end{equation}
where $l_L$ is the lepton doublet, the Majorana mass $M$ is chosen to be real and 
we have assumed that a single $\nu_R$ species dominates.
The quartic and trilinear Higgs-inflaton couplings are generated at 1 loop and the result is divergent. In other words, such couplings are required by renormalizability of the model. 
As the renormalization condition, we take $\lhf(M_{\rm Pl})=0$,
$\sigma_{h\phi}(M_{\rm Pl})=0$ such that at the inflationary scale 
the couplings are generated by loop effects.  
In the
leading--log approximation, we find
\begin{eqnarray}
\lambda_{h\phi}&\simeq& { \vert \lambda_\nu y_\nu \vert^2 \over 2 \pi^2} \ln {M_{\rm Pl}\over \mu} \;, 
\nonumber\\
\sigma_{h\phi}&\simeq& -{ M \vert y_\nu \vert^2 {\rm Re} \lambda_\nu \over 2 \pi^2} \ln {M_{\rm Pl}\over 
\mu} \;,
\end{eqnarray}
where $\mu$ is the relevant energy scale.
In what follows, we assume real couplings and
take $\mu \sim m$ since this is the typical momentum of the Higgs quanta towards the end of the resonance.\footnote{Choosing a higher $\mu$  
would result in slightly looser bounds.} In any case, the dependence 
on $\mu$ is only logarithmic.

The value of $\lambda_\nu$ is constrained by inflationary dynamics.
In order not to spoil flatness of the inflaton potential,  the coupling must satisfy $\lambda_\nu < 10^{-3}$ \cite{Gross:2015bea}.
Taking $\lambda_{h\phi}< 3\times 10^{-8}$ and $|\sigma_{h\phi}| < 10^8$ GeV (see fig.~\ref{fig:destb}), we find the following bounds 
on the neutrino Yukawa coupling and the Majorana mass,
\begin{eqnarray}
&& y_\nu < 0.2\nonumber \;, \\
&& M<  4 \times 10^{12} \; {\rm GeV} \;.
\end{eqnarray}
Although these constraints are not particularly strong, they are non--trivial. In particular, they imply that the neutrino Yukawa coupling cannot be order one.

\subsection{Reheating via Non--renormalizable Operators}

A common $^{}$approach to $^{}$reheating is to assume the presence 
of $^{}$non--renormalizable operators that $^{}$couple the inflaton to the SM fields. Let us consider 
a $^{}$representative example of the $^{}$following $^{}$operators,
\begin{equation}
O_1 = {1\over \Lambda_1} \phi\; \bar q_L \! \cdot \!  H^* \, t_R ~~,~~
O_2 = {1\over \Lambda_2} \phi\; G_{\mu\nu} G^{\mu\nu} ~, 
\end{equation} 
where $\Lambda_{1,2}$ are some scales, $G_{\mu\nu}$ is the gluon field strength and $q_L,t_R$ are 
the third generation quarks. These $^{}$couplings allow for a $^{}$direct $^{}$decay of the $^{}$inflaton into the SM particles.
It is $^{}$again clear that a Higgs--inflaton $^{}$interaction is induced $^{}$radiatively. In order to $^{}$calculate the 1--loop 
couplings $^{}$reliably, one $^{}$needs to $^{}$complete the model in the $^{}$ultraviolet (UV). The simplest $^{}$possibility to
obtain an $^{}$effective dim-5 $^{}$operator is to $^{}$integrate out a $^{}$heavy fermion. Therefore, we $^{}$introduce $^{}$vector--like 
quarks $Q_L, Q_R$ with the tree $^{}$level $^{}$interactions
\begin{equation}
-\Delta {\cal L} = {y_Q} \; \bar q_L \! \cdot \!  H^* \, Q_R + {\lambda_Q } \; \phi \;
\bar Q_L t_R + {\cal M} \; \bar Q_L Q_R + {\rm h.c.}~, 
\end{equation}
where the heavy $^{}$quarks have the $^{}$quantum numbers of the $^{}$right--handed top $t_R$,
their mass 
${\cal M}$ is $^{}$taken to be 
above the $^{}$inflaton mass $^{}$scale and the $^{}$couplings to the third $^{}$generation are assumed to $^{}$dominate.
One then$^{}$ finds that $O_1$ appears at $^{}$tree level $^{}$with
$1/\Lambda_1 = y_Q \lambda_Q/{\cal M}$, whereas $O_2$ $^{}$appears only at 2 $^{}$loops 
with $1/\Lambda_2 \sim y_Q \lambda_Q y_t \alpha_s/(64 \pi^3 {\cal M})$ and
can be $^{}$neglected. Using the $^{}$renormalization $^{}$condition that the $^{}$relevant
couplings $^{}$vanish at the Planck $^{}$scale and the fact that the heavy quarks $^{}$contribute only at scales above ${\cal M}$, we get in the $^{}$leading--log
$^{}$approximation
\begin{eqnarray}
\lambda_{h\phi}&\simeq& { 3 \vert \lambda_Q y_t \vert^2 \over 2 \pi^2} \ln {M_{\rm Pl}\over {\cal M}} \;, \nonumber\\
\sigma_{h\phi}&\simeq& -{ 3 {\cal M} \;{\rm Re} (\lambda_Q y_Q y_t) \over 2 \pi^2} \ln {M_{\rm Pl}\over {\cal M}} \;,
\end{eqnarray}
where $y_t$ is the top Yukawa $^{}$coupling and we $^{}$assume
${\cal M}\ll M_{\rm Pl}$. 
As in the $^{}$previous example, one of the couplings is constrained
by the $^{}$inflationary
dynamics,  
$|\lambda_Q| < 2 \times 10^{-3}/ (\ln \;M_{\rm Pl}/{\cal M} )^{1/4}$
 \cite{Gross:2015bea},
since it $^{}$generates a correction to the inflaton potential.
The heavy quark mass must be well below the Planck scale, 
$\mathcal M \ll M_{\rm Pl}$, and the bound on $\lambda_Q$ depends
very weakly on $\mathcal M$ in the allowed range. Therefore, in 
practice one may take $|\lambda_Q |<  10^{-3}$.
Our results $\lambda_{h\phi} < 3\times 10^{-8}$,
$|\sigma_{h\phi}|< 10^8$ GeV   lead to  a stronger bound. For real couplings, we get
 \begin{eqnarray}
&& |\lambda_Q| < 4 \times 10^{-4}\nonumber \;, \\
&& |y_Q| <0.02 \;,
\end{eqnarray}
where in the second inequality we took ${\cal M}\sim m$ to obtain
the most conservative bound and $y_t({\cal M})\sim 1/2$.
This implies, in particular, that the minimal value of 
the suppression scale $\Lambda_1=
{\cal M}/|\lambda_Q y_Q|$ is around the Planck scale and the maximal
reheating temperature is of order $10^9$ GeV (see \cite{Gross:2015bea} for details).

\section{Conclusions}

This work is devoted to an in--depth analysis of the Higgs--inflaton coupling effects in the reheating epoch. We have focussed in particular on the preheating stage when the parametric and tachyonic resonances are active.  Our framework includes both the quartic and trilinear couplings since these
are present simultaneously in realistic models. 
The resulting mixed parametric--tachyonic resonance is described by the Whittaker--Hill equation.
While inheriting  certain features of the two resonances, it 
  brings in new effects which require a thorough investigation.

Within this framework, we have analyzed the issue of electroweak vacuum 
stability during the preheating epoch assuming that the Higgs 
self--coupling turns negative at high energies.
Even though the Higgs--inflaton
couplings can stabilize the system during inflation, resonant Higgs production thereafter can lead to vacuum destabilization.
The relevant quartic and trilinear Higgs--inflaton couplings are 
generated by the renormalization group equations in realistic models,
and even their tiny values make a difference. Using both analytical
methods and lattice simulations in a representative  large field ($\phi^2$) inflation model,  we obtain upper bounds on the couplings from vacuum stability during preheating.  
These allow for   a range of couplings, roughly 
$10^{-10} < \lambda_{h \phi} < 10^{-8} $ and $|\sigma_{h\phi}|< 10^8$ GeV,
 which ensure 
 stability both during inflation and preheating.

 Our analysis is limited to the timescale of the mixed resonance. This leaves out the  
issues of the late--time behavior of the Higgs fluctuations which
 can further limit
the allowed range for the couplings. The required analysis is highly involved and we leave it for future work.

\section*{Acknowledgments}

MK is supported by the Academy of Finland project 278722 and during the initial stages of this work was supported by JSPS as an International Research Fellow of the Japan Society for the Promotion of Science. O.L. and M.Z. acknowledge support from the Academy of Finland project ``The Higgs and the Universe''.

\appendix

\section{The Whittaker-Hill Equation\label{sec:WH-Eqn}}

\subsection{Computation of the Floquet Exponent}

The Whittaker-Hill equation is given by  
\begin{eqnarray}
\left[\frac{\mathrm{d}^{2}}{\mathrm{d}z^{2}}+2p\cos\left(2z\right)+2q\cos\left(4z\right)\right]X & = & -AX,\label{WH}
\end{eqnarray}
where \mk{a constant $A$}
%\begin{equation}
%A\left(k\right)\equiv\left(2\frac{k}{m}\right)^{2}+2q\label{Ac-def}
%\end{equation}
can be thought of as an eigenvalue of the differential operator on
the LHS. The analysis of this equation can be found in 
\cite{Lachapelle(2009)preh,Whittaker(1996)preh,Roncaratti(2010)preh,Possa(2016)preh}.

According to the Floquet  theorem, the solution can be written as  a  series of the form
\begin{equation}
X\left(z\right)=\mathrm{e}^{\mu z}\sum_{n=-\infty}^{\infty}c_{2n}\mathrm{e}^{2inz}.
\end{equation}
Plugging this Ansatz into the above equation, we obtain a recursive relation
\begin{equation}
\gamma_{2n}\left(c_{2\left(n-1\right)}+c_{2\left(n+1\right)}\right)+c_{2n}+\xi_{2n}\left(c_{2\left(n-2\right)}+c_{2\left(n+2\right)}\right)=0,
\end{equation}
where
\begin{eqnarray}
\gamma_{2n}\equiv\frac{p}{A-\left(i\mu-2n\right)^{2}} & \quad\mathrm{and}\quad & \xi_{2n}\equiv\frac{q}{A-\left(i\mu-2n\right)^{2}}.\label{gamma-def}
\end{eqnarray}
For given values of \mk{$A$}, $q$ and $p$ we can find the Floquet
characteristic exponent $\mu$ by solving for the roots of the determinant
\begin{equation}
\Delta\left(i\mu\right)=\left|\begin{array}{ccccccccc}
\ddots\\
 & \xi_{-2} & \gamma_{-2} & 1 & \gamma_{-2} & \xi_{-2} & 0 & 0\\
 & 0 & \xi_{0} & \gamma_{0} & 1 & \gamma_{0} & \xi_{0} & 0\\
 & 0 & 0 & \xi_{2} & \gamma_{2} & 1 & \gamma_{2} & \xi_{2}\\
 &  &  &  &  &  &  &  & \ddots
\end{array}\right|=0
\end{equation}
\mk{It is possible to prove} (see, e.g., refs~\cite{Whittaker(1996)preh,Lachapelle(2009)preh})
that this determinant can be written in a compact form as
\begin{equation}
\sin^{2}\left(i\mu\frac{\pi}{2}\right)=\Delta\left(0\right)\sin^{2}\left(\sqrt{A}\frac{\pi}{2}\right).
\end{equation}
From this equation we can easily find
\begin{equation}
\mu=-\frac{i}{\pi}\arccos\left[1+\Delta\left(0\right)\left(\cos\left(\sqrt{A}\pi\right)-1\right)\right].\label{mu-sol}
\end{equation}
The advantage of this representation of solutions is that it can be
evaluated numerically very efficiently. Indeed, as one can see from
the definitions of $\gamma_{2n}$ and $\xi_{2n}$ in eqs.~(\ref{gamma-def}),
the off-diagonal elements of $\Delta\left(i\mu\right)$ decrease as
$\propto n^{-2}$ as they depart from the center of the matrix.

\subsection{Boundary Between Stability and Instability Regions}

The stability of the solution in eq.~(\ref{WH-gensol}) is determined
by the characteristic exponent $\mu$. In general, $\mu$ is a complex
number $\mu=\alpha+i\beta$. If the real part of $\mu$ is non-zero,
that is $\alpha\ne0$, the given solution is unstable. For stable
solutions $\alpha=0$ and their periodicity is determined by the value
of the imaginary part $\beta$. If $\beta$ is a rational fraction,
the solution is periodic, while for irrational $\beta$ the solution
is non-periodic. Particularly interesting are the cases where $\beta$
is an integer. If $\beta=2l$, where $l\in\mathbb{Z}$, then solutions
are either even or odd periodic functions with a period $\pi$. For
$\beta=2l+1$ those solutions are even or odd periodic functions with
a period $2\pi$. These solutions of   period $\pi$ and $2\pi$
lie on the boundary between the regions where the families of stable
and unstable solutions reside, that is, the so called stability and instability
regions  in $\left(A,q,p\right)$ space. To find the equations for
these boundary surfaces, we can use the following Ans\"atze
\begin{eqnarray}
y_{1}\left(z\right) & = & \sum_{n=0}^{\infty}C_{2n}\cos\left(2nz\right)~,\label{y1-def}\\
y_{2}\left(z\right) & = & \sum_{n=0}^{\infty}S_{2n+1}\sin\left(\left(2n+1\right)z\right)~,\\
y_{3}\left(z\right) & = & \sum_{n=0}^{\infty}C_{2n+1}\cos\left(\left(2n+1\right)z\right)~,\\
y_{4}\left(z\right) & = & \sum_{n=0}^{\infty}S_{2n+2}\sin\left(\left(2n+2\right)z\right)~.\label{y4-def}
\end{eqnarray}
The   function  $y_{1}\left(z\right)$
describes   even solutions of period $\pi$; $y_{2}\left(z\right)$
is an odd function of period $2\pi$; $y_{3}\left(z\right)$ 
is an even function  of period $2\pi$ and $y_{4}\left(z\right)$  is an odd function
of period $\pi$. To find the coefficients, one plugs these functions
back into the Whittaker-Hill equation. This gives us four
recursive relations which can be written in a matrix form,
\begin{equation}
M_{J}C^{J}=A_{J}C^{J},
\end{equation}
where no summation over $J$ is implied; $C^{J}$ stands
for $C^{1}=\left(C_{0},C_{2},C_{4},\ldots\right)^{\mathrm{T}}$, $C^{2}=\left(S_{1,}S_{3},\ldots\right)^{\mathrm{T}}$,
 etc. and  $A_{J}$ represent eigenvalues $A_{IJ}\left(p,q\right)$
of an infinite square matrix $M_{J}$, where $I=0,1,\ldots,\infty$.
These eigenvalues define the boundaries between the stability and
instability regions.

\begin{figure}
\centering{}\includegraphics[scale=0.37]{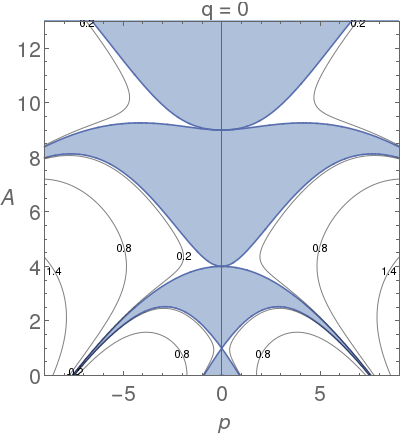}~\includegraphics[scale=0.37]{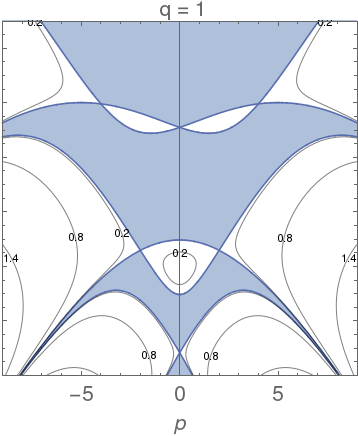}~\includegraphics[scale=0.37]{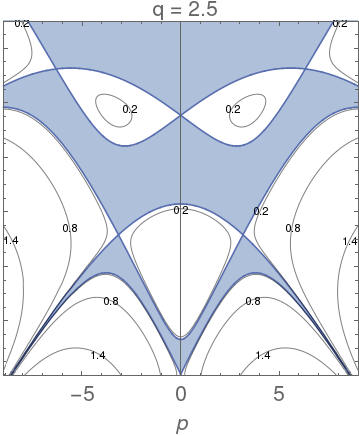}\caption{\label{fig:WH-stb-istb}Stability (shaded) and instability
(white) regions of the Whittaker-Hill equation
for several values of $q$. The solid curves 
are contours of constant $|$Re$\mu|$. 
The leftmost panel ($q=0$) is the  stability chart
of the Mathieu equation.}
\end{figure}

To find $A_{IJ}\left(p,q\right)$, let us compute
the four matrices $M_{J}$ explicitly.
Plugging eq.~(\ref{y1-def}) into eq.~(\ref{WH}), we find
\begin{equation}
M_{1}=\left(\begin{array}{ccccc}
0 & p & q & 0\\
2p & q-4 & p & q\\
2q & p & -16 & p\\
0 & q & p & -4n^{2}\\
 &  &  &  & \ddots
\end{array}\right),\label{M1}
\end{equation}
where the first row corresponds to $n=0$. Similarly, one   obtains
the other three matrices,
\begin{equation}
M_{2}=\left(\begin{array}{cccccc}
-p-1 & p-q & q & 0\\
p-q & -9 & p & q\\
q & p & -25 & p\\
0 & q & p & -\left(2n+1\right)^{2}\\
 &  &  &  & \ddots
\end{array}\right),
\end{equation}
 
\begin{equation}
M_{3}=\left(\begin{array}{ccccc}
p-1 & p+q & q & 0\\
p+q & -9 & p & q\\
q & p & -25 & p\\
0 & q & p & -\left(2n+1\right)^{2}\\
 &  &  &  & \ddots
\end{array}\right),
\end{equation}
\begin{equation}
M_{4}=\left(\begin{array}{ccccc}
-q-4 & p & q & 0\\
p & -16 & p & q\\
q & p & -4\left(n+1\right)^{2} & p\\
 &  &  &  & \ddots
\end{array}\right).\label{M4}
\end{equation}
We compute the eigenvalues of these matrices numerically by truncating
them at some high value of $n$. Some solutions of eq.~(\ref{mu-sol})
and eigenvalues of these matrices are shown in figure~\ref{fig:WH-stb-istb}.
The leftmost panel ($q=0$) can be recognized as the familiar stability chart
of the Mathieu equation.

%\bibliographystyle{JHEP}
%\bibliography{HiggsBib}

\providecommand{\href}[2]{#2}\begingroup\raggedright\endgroup

\end{document}